\documentclass[aps,superscriptaddress,twocolumn,showpacs]{revtex4-1}
\usepackage{subfig}
\usepackage{color}
\usepackage[mathscr]{euscript}
\usepackage{graphicx}
\usepackage{amsmath}
\usepackage{array}
\usepackage{multirow}
\newcommand{\ra}[1]{\renewcommand{\arraystretch}{#1}}

\newcommand{\Ln}{\mathcal{L}}

\begin{document}
\title{Parameter estimation of a nonlinear magnetic universe from observations}
\author{Ariadna Montiel$^{1}$, Nora Bret\'on$^{1}$ and Vincenzo Salzano$^{2}$}
\affiliation{
$^{}$ Dpto. de F\'{\i}sica, Centro de Investigaci\'on y de Estudios Avanzados 
del I. P. N., Apdo. 14-740, D.F., Mexico. \\
$^{2}$ Fisika Teorikoaren eta Zientziaren Historia Saila, Zientzia eta Teknologia Fakultatea, Euskal Herriko Unibertsitatea UPV/EHU, 644 Posta Kutxatila, 48080 Bilbao, Spain. }

\begin{abstract}
The cosmological model consisting of a nonlinear magnetic field obeying the Lagrangian $\Ln= \gamma F^{\alpha}$, $F$ being the electromagnetic invariant, coupled to a Robertson-Walker geometry is tested with observational data of Type Ia Supernovae, Long Gamma-Ray Bursts and Hubble parameter measurements. 
The statistical analysis show that the inclusion of nonlinear electromagnetic matter is enough to produce the observed accelerated expansion, with not need of including a dark energy component. The electromagnetic matter with abundance $\Omega_B$, gives as best fit from the combination of all observational data sets  $\Omega_B=0.562^{+0.037}_{-0.038}$ for the scenario in which $\alpha=-1$,  $\Omega_B=0.654^{+0.040}_{-0.040}$ for the scenario with $\alpha=-1/4$ and $\Omega_B=0.683^{+0.039}_{-0.043}$ for the one with $\alpha=-1/8$. These results indicate that nonlinear electromagnetic matter could play the role of dark energy, with the theoretical advantage of being a mensurable field.
\end{abstract}

\maketitle

\section{Introduction}

According to Einstein's equations and assuming a Robertson-Walker (RW) geometry, the currently inferred accelerated expansion of the universe is attributed to a kind of repulsive gravity that makes fall apart spacetime. Such expansion is possible if the dominant component of the universe, the so called dark energy (DE), acts with a negative pressure that overcomes the attractive effect of ordinary matter; its corresponding energy density $\rho$ and pressure $p$ should be such that $\rho + 3 p <0$, in order to produce the mentioned acceleration.

It has been shown that the effect of coupling nonlinear electrodynamics to gravity
produces negative pressures that in turn accelerate the expansion \cite{Novello2,Novello3,Vollick2008,Labun2010}. In \cite{Dyadichev2002} cosmological models involving homogeneous and isotropic Yang-Mills fields were proposed as an alternative to scalar models of cosmic acceleration; while in \cite{Elizalde2003} a quantum condensate is considered as driven the accelerated expansion. In \cite{Jimenez:2008au} it is shown that a vector-tensor theory consisting of gauge fields coupled to gravity could be the origin of the accelerated expansion of the Universe. In \cite{Jimenez:2008nm} it is pointed out that an effective cosmological constant may arise from an electromagnetic mode or degree of freedom, considering that the electromagnetic field contains an additional (scalar) polarization, such that  quantum fluctuations of the energy density get frozen on cosmological scales giving rise to an effective cosmological constant. In \cite{Jimenez:2008er} a timelike electromagnetic field on cosmological scales generates an effective cosmological constant; this field could be originated in primordial electromagnetic quantum fluctuations producing during the inflationary epoch.
These models open the possibility that DE originates in properties of ponderable fields and matter.

Unlike early universes where high energies justify the appearance of nonlinear electromagnetic effects, in late epochs,  the reason to invoke nonlinear electromagnetic behaviour may be different:
it can be implemented as a phenomenological approach \cite{Medeiros2012}, in which the cosmic substratum is modeled as a material media with electric permeability and magnetic susceptibility that depend  in nonlinear way on the fields \cite{Pleban}. 
Another argument relies in the view that General Relativity is a low energy quantum effective field theory of gravity, provided that the Einstein-Hilbert classical action is augmented by the additional terms required by the trace anomaly characteristic of nonlinear electrodynamics \cite{Mottola}.

Assuming that the cosmological background affects the transmission of light signals,
there is another approach that considers nonlinear behaviour in the propagation of light, similar to light traveling in non vacuum spacetime \cite{Mosquera2}. This approach has its basis in the fact that the nonlinear electromagnetic Born-Infeld equations are  of the same form than Maxwell's for a material media with the difference that the electric permeability and magnetic susceptibility are functions of the field strengths \cite{BI}. 

A technical problem arises in the coupling of an electromagnetic field  
to an isotropic geometry, as the electromagnetic field defines preferred directions, so an isotropization process of the energy-momentum tensor should be adopted. To this end several proposals have come up: one of them is to take a spatial average in the electromagnetic field, \cite{Tolman,Vollick2008,Elizalde2003,Novello2,Novello3},  
Alternatively, it has been considered a vector triplet compatible with space homogeneity and isotropy of RW \cite{ArmendarizP}.
This is a set of three equal length vectors that point in three mutually orthogonal spatial directions. While the triad guarantees the isotropy of the background, it does not automatically imply the isotropy of its perturbations that are necessary to model some observed anomalies in the CMB radiation. In fact the cosmic triad can be realized with a classical SU(2) vector field configuration \cite{ArmendarizP,Dyadichev2002}.

The purpose of this work is to investigate to what extent nonlinear magnetic matter can be considered as source of the present cosmic acceleration as an alternative to the DE component. 
We shall consider a phenomenological model with a nonlinear magnetic field, proposed in \cite{Novello3}, associated to the nonlinear Lagrangian $\Ln=\gamma F^{\alpha}$, where $\gamma$ and $\alpha$ are two constants to be adjusted from  observations. We perform a $\chi^2$ statistical analysis by using a Markov Chain Monte Carlo (MCMC) code; we probe the model with Type Ia Supernovae (SNe Ia), Long Gamma-Ray Bursts (LGRBs) and observational Hubble data (OHD). We analyze three cases, namely, $\alpha=-1$, $\alpha=-1/4$ and $\alpha=-1/8$. We could possibly think of considering a time dependent $\alpha$, which in turn, would lead to a time dependent equation of state (EoS) parameter, $w(z)$, however, a constant $w$ has the great advantage of simplicity and that is why we performed the analysis with fixed $\alpha$.  
In all cases, we obtain good best fits without introducing the DE component. 
  
The paper is organized as follows. In Section 2 we address the coupling of nonlinear electrodynamics (NLED) to a RW geometry. In Section 3, theoretical details of the nonlinear magnetic universe are given. In Section 4 the observational data samples and the statistical method used are presented. In Section 5 the obtained constraints and best fits are discussed, and finally the last section is for concluding remarks. 

\section{Coupling nonlinear electrodynamics to RW}
 
The four-dimensional Einstein-Hilbert action of gravity coupled to NLED is given by
\begin{equation}
S= \frac{1}{16 \pi }\int{ \sqrt{-g} d^4x \left( - {R}  + \Ln(F,G) \right)},
\label{Eq:action}
\end{equation}
where $R$ is the Ricci scalar and $\Ln(F,G)$ is the electromagnetic Lagrangian that depends on the electromagnetic invariants $F=F_{\mu \nu}F^{\mu \nu}= 2(B^2-E^2)$ and $G=(\sqrt{-g}/2) \epsilon_{\mu \nu \rho \sigma}F^{\rho \sigma}F^{\mu \nu}=4 E \cdot B$, where  $\epsilon_{\mu \nu \rho \sigma}$ is the Levi-Civita symbol; $E$ and $B$ are the electric field and magnetic induction, respectively.

As we mentioned before, several mechanisms to isotropize the electromagnetic energy-momentum tensor have been proposed so far. Despite its intrinsical anisotropic evolution, in \cite{Cembranos:2012kk} it has been shown that the average energy-momentum tensor associated to rapid evolving vector field is isotropic under very general and natural conditions. As it is not clear if this criteria would apply also for nonlinear electromagnetic fields, 
we shall assume the spatial average proposed by Tolman and Ehrenfest (1933) \cite{Tolman}.
The resulting isotropic energy-momentum tensor, with energy density $\rho=T_{0}^{0}$ and pressure $p=- T_{i}^{i}/3,~~ i=1,2,3$, is given by 
\begin{eqnarray} 
T_{\mu \nu}&=&-4 \Ln_F F_{\mu \cdot}^{\alpha}F_{\alpha \nu}+(G\Ln_G -\Ln) g_{\mu \nu}  \nonumber \\
                   &=&(\rho + p) u_{\mu} u_{\nu}-pg_{\mu \nu}, \nonumber\\  
\rho&=& -\Ln+G\Ln_G-4E^2\Ln_F,\nonumber\\
p&=& \Ln-G\Ln_G+\frac{4}{3}(E^2-2B^2)\Ln_F,
\label{Eq:EnergyDensity}
\end{eqnarray}
where $\Ln_X=d\Ln/dX$.

In this work we shall consider a Lagrangian consisting of the Maxwell term and the nonlinear term,

\begin{equation}
\Ln(F)= - \frac{F}{4}+{\gamma}{F^{\alpha}}.
\label{nled_lagr}
\end{equation}
 
Since we
are interested in the late epoch of the Universe and in reproducing the observed accelerated expansion with the
nonlinear term, in the forthcoming analysis we shall neglect the linear term; as it is related to the CMB radiation, whose order of magnitude is $\Omega_{rad}=2.47 \times 10^{-5} h^{-2}$, smaller than the dark energy density by far. We shall address the cases $\alpha=-1$, $\alpha=-1/4$ and $\alpha=-1/8$ successively.

\section{Nonlinear magnetic universe}
  
The scenario in which $E^2=0$, called \textit{magnetic universe}, is the relevant one in cosmology \cite{Novello3,Novello:2005bj,Lemoine:1995vj}. Cosmological magnetic universes have been explored before, for instance in \cite{Novello:2008xp} a cyclic magnetic cosmological toy model was introduced; from this model arose a complete cyclic scenario
consisting of five noninteracting perfect fluids that evolve independently and whose parameters 
were adjusted using SNe Ia and CMB in  \cite{Medeiros2012}. The one regarding the accelerated expansion arises from a term in the Lagrangian of the form
$\Ln(F) \propto {- \mu^2}{F^{-1}}$; since a bouncing is a possibility in this model, it was not
considered that $ \sum \Omega_i=1$. A similar nonlinear magnetic scenario was considered in higher dimensions in \cite{Chayan2013} and some parameters were constrained.

In this paper we study the nonlinear magnetic scenario described by the Lagrangian $\Ln=\gamma F^{\alpha}$ with $F=2B^2$. 
This Lagrangian resembles several noteworthy (purely magnetic) ones, for instance, Born-Infeld Lagrangian is obtained with $\alpha=1/2$; if $\alpha=2$, it has the form of the Euler-Heisenberg Lagrangian \cite{EulerHeis}, the Abelian Pagels-Tomboulis one \cite{Pagels:1978dd} is also included. The case $\alpha=-1$ has been studied previously in \cite{Novello4}, but it has not been observationally tested.

Before procceding to the analysis, a comment on the hyperbolicity of the equations derived from Lagrangians
of the kind of Eq.(\ref{nled_lagr}) is in order.
In \cite{EspositoFarese:2009aj}, it is shown that for a vector field with an action of the form

\begin{equation}
S= - \int{d^4x \left[ f(F)+V(A^2) \right]},
\label{Eq:action2}
\end{equation}

where $A^2= A_{\mu}A^{\mu}$, the well-posedness of Chauchy problem breaks down somewhere in the allowed phase space.
However in \cite{Golovnev:2013gpa} the problem was revisited and it was proved that hyperbolicity violations do not appear around homogeneous field configurations necessarily. The authors considered spatial homogeneous fields $A^{\mu}(t)$ in FRW spacetimes and derived hyperbolicity criteria based on the signs of the derivatives  of $f(F)$; a detailed analysis considering the behaviour of $B/a$ is needed in order to apply such criteria; in anycase the authors mentioned that a fine tunning is always possible to obtain well behaved equations.

The energy density and effective pressure, Eq. (\ref{Eq:EnergyDensity}), derived from the nonlinear term
in Eq. (\ref{nled_lagr}) are

\begin{equation}
\rho_B=-\Ln, \qquad p_B= \Ln -\frac{4}{3}F \Ln_F.
\label{Eq:rhoMU}
\end{equation}

The corresponding field equations are derived from the action, Eq. (\ref{Eq:action}), by performing variations with respect to the metric $g_{\mu \nu}$. For the RW metric with a perfect fluid, the Friedmann equations are
\begin{eqnarray} 
H^2&\equiv & \left({\frac{\dot{a}}{a}}\right)^2= \frac{\rho}{3}, \nonumber\\
3 \frac{\ddot{a}}{a}&=& - \frac{1}{2} (\rho+3p), \nonumber\\
\label{Eq:FriedmannEqs}
\end{eqnarray}
where $a$ is the scale factor, $H$ is the Hubble parameter and the overdot means derivative with respect to the cosmic time $t$. Here we have set $c=1$. 

From the second Friedmann equation, the condition to produce accelerated expansion is that $(\rho+3p)<0$. For the magnetic universe this condition can be written, using Eq. (\ref{Eq:rhoMU}), as
\begin{equation}
\rho+3p=2\Ln-4F\Ln_F<0, \quad \Longleftrightarrow \quad \Ln<2F\Ln_F.
\label{Eq.5}
\end{equation}
In particular, for the Lagrangian of the form $\Ln= \gamma F^{\alpha}$, with $\gamma<0$, the accelerated expansion condition
is fulfilled if $\alpha <1/2$. 
 
From the energy conservation law, $\dot{\rho}+3 H(\rho+p)=0$, the scaling between the electromagnetic field and the scale factor, $F=({\rm const}) a^{-4}$ can be derived, see Appendix A for details.
Consequently, the magnetic field scales as $B \sim a^{-2}$. Notice that this result does not depend on the particular analytic form of $\Ln(F)$. On the other side, for the Lagrangian $\Ln= \gamma F^{\alpha}$ knowing that $Fa^4=$ const, it can be shown that $\Ln a^{4 \alpha}=$ const and then the equations can be integrated to obtain $a(t)$, see Appendix B.

The energy density for the nonlinear magnetic component is obtained by using $B \sim a^{-2}$, such that
\begin{equation}
\rho_{B}= -2^{\alpha} B_0^{2\alpha} \gamma  a^{-4 \alpha},
\label{Eq:rhoB}
\end{equation}
where $B_0$ is an integration constant, $B=B_0a^{-2}$
and $\gamma$ must be negative in order to have a positive energy density, $\rho_B>0$.

\begin{figure}
  \centering
  \includegraphics[width=0.495\textwidth]{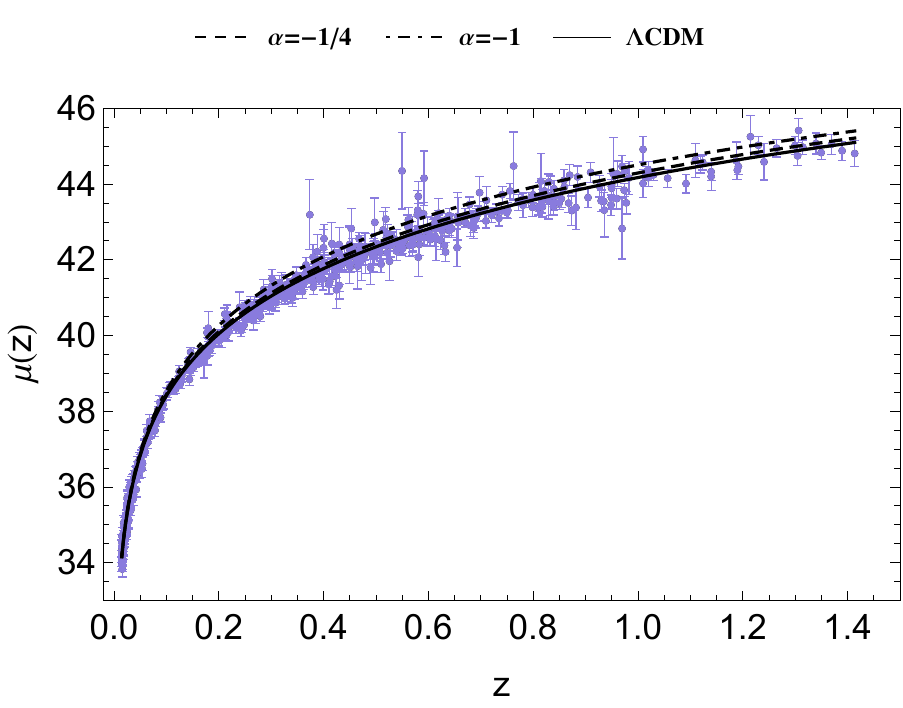}
  \caption{Hubble Diagram from Type Ia Supernovae (Union2.1 compilation) and theoretical prediction from $\Lambda$CDM model (solid line) and the nonlinear magnetic universes with $\alpha=-1$ (dot-dashed line) and $\alpha=-1/4$ (dashed line), in which we have assumed $\Omega_m$ and $H_0$ from the Planck results \cite{PlanckXXVI}.}
  \label{Fig:Mu}
\end{figure}

We will assume a two-component universe made of dust matter, $\rho_m \propto a^{-3}$, and the nonlinear magnetic component characterized by $\rho_B \propto a^{-4\alpha}$ with equation of state (EoS) $p=w \rho_B=(4\alpha/3 -1) \rho_B$. Note that the $\Lambda$CDM model is recovered by taking $\alpha=0$, however, since we do not know a priori what is the true value of $\alpha$,  we test for different values of $\alpha$ (see Fig.\ref{Fig:Mu}).
Some authors have also suggested a non-constant EoS-parameter, derived from a variation of the cosmological constant with an energy scale associated to the renormalization group running; such scale can be identified with the Hubble parameter and the cosmological term could inherit that time-dependence 
through its primary scale evolution with the renormalization scale parameter. A dynamical EoS for the dark energy implies that the EoS-parameter $w$ should be evolving with the redshift, that usually is interpreted as dark energy with a scalar field origin \cite{Sola:2005et}. 

The Hubble parameter in terms of the redshift and the fractional energy densities then reads,
\begin{equation}
\frac{H^2(z)}{H^2_0}=\Omega_{m}(1+z)^3 + \Omega_{B}(1+z)^{4\alpha},
\label{Eq:H}
\end{equation}  
with $\Omega_m={\rho^0_m}/{\rho_{c,0}}$ and $\Omega_B={- \gamma 2^{\alpha} B_0^{2 \alpha}}/{\rho_{c,0}}$.
The constant $\gamma$ should be adjusted in order to have energy density units in the Lagrangian $\gamma F^{\alpha}$. Note that by taking appropriate values of $\alpha$, Eq. (\ref{Eq:H}) leads to a phantom DE scenario \cite{Phantom}.

Regarding the kinematical approach, in which the deceleration parameter $q$ is parameterized as a function of the redshift $z$,  it is straightforward to obtain $q(z)$ as function of the free parameters of the model using the EoS, $w=4\alpha/3-1$, and the Hubble parameter $H(z)$, Eq. (\ref{Eq:H}), as follows:
\begin{equation}
q(z)=\frac{3}{2}\left(1-\frac{\Omega_m\left(1+z\right)^3}{(H/H_0)^2}\right)w+\frac{1}{2},
\end{equation}
that explicitly is,
\begin{equation}
q(z)= \frac{1}{2} \left[ \frac{2(2\alpha - 1)\Omega_B(1+z)^{4\alpha}+\Omega_m(1+z)^3}{\Omega_m(1+z)^3+\Omega_B(1+z)^{4\alpha}}\right].
\label{Eq:q}
\end{equation}

At the present time, $z=0$, Eqs. (\ref{Eq:H}) and (\ref{Eq:q}) read
\begin{eqnarray}
1 & = & \Omega_m + \Omega_B, \label{Eq:norm}\\
q_0 & = & \frac{1}{2} \left[ \frac{2(2\alpha-1)\Omega_B+ \Omega_m}{\Omega_m + \Omega_B} \right]. \label{Eq:q0}
\end{eqnarray}

Eq. (\ref{Eq:norm}) resembles the standard $\Lambda$CDM model, with $ \Omega_{\Lambda} \mapsto \Omega_B$; moreover, by using this Eq.  (\ref{Eq:norm}), we can reduce the parameter-dimension of the problem to only two free parameters, namely, $\Omega_m$ and $H_0$,  when we use the observational Hubble data as well as for the combination of all observational data sets. In our analysis we shall use the dimensionless Hubble constant $h$ instead $H_0$, they are related through $H_0= 100h$ km s$^{-1}$ Mpc$^{-1}$. Furthermore, Eq.  (\ref{Eq:q0}) indicates that the acceleration of the universe (i.e. $q_0 < 0$) in the nonlinear magnetic universe can arise from  $\alpha$ fulfilling 
\begin{equation}
\alpha<\frac{1}{2}-\frac{\Omega_m}{4\Omega_B}.
\label{Eq:alpha}
\end{equation}

\section{Observational data sets and Statistics}

\subsection{Type Ia Supernovae (SNe Ia)}

To test the nonlinear magnetic scenarios against cosmological observations, we first consider the updated Union2.1 compilation of 580 SNe Ia reported by the Supernova Cosmology Project (SCP) \cite{Union21}. 

The comparison with SNe Ia data is made via the standard $\chi^2$ statistics given by
\begin{equation}
\chi^2_{SNe Ia}= \mathrm{\Delta} \mathrm{F} \cdot \mathrm{C^{-1}} \cdot \mathrm{\Delta} \mathrm{F},
\label{Eq:CSNIa}
\end{equation}
where $\mathrm{C}$ is the covariance matrix and $\mathrm{\Delta} \mathrm{F}=\mathrm{F_{th}}-\mathrm{F_{obs}}$ is the vector of the differences between the observed and theoretical value of the quantity $\mathrm{F}$. For Union2.1, $\mathrm{C}$ captures all identified systematic errors besides to the statistic errors of the SNe Ia data and $\mathrm{F}$ corresponds to the distance modulus 
\begin{equation}
\mu(z,\theta)= 5 \log_{10} \left[d_L(z,\theta) \right] + \mu_0,
\label{Eq:muSN}
\end{equation}
where $d_L(z,\theta) $ is the dimensionless luminosity distance given by 
\begin{equation}
d_L(z,\theta)= (1+z) \int_0^z \frac{dz'}{E(z',\theta)},
\end{equation}
with $E(z,\theta)=H(z,\theta)/H_0$ the dimensionless Hubble function, $H_0$ the Hubble constant and $\theta$ the free parameters of the cosmological model. 

In Eq. (\ref{Eq:muSN}) $\mu_0$ is a nuisance parameter that depends on both the absolute magnitude of a fiducial SN Ia and the Hubble constant. In this work, we marginalize the $\chi^2_{SNe Ia}$ over $\mu_0$.

\subsection{Observational Hubble Data (OHD)}

The observational Hubble parameter (OHD), compared with other observational techniques, provides a direct measurement of the Hubble parameter, and not of its integral, unlike SNeIa or angular/angle-averaged BAO. Thus, this independent dataset can help break the parameter degeneracies and shed light on the cosmological scenarios and in particular, on the nonlinear magnetic scenarios. 

In this work, we use 18 data points from differential evolution of passively evolving early-type galaxies in the redshift range $0<z<1.75$ recently updated in \cite{Jimenez12} but first reported in \cite{Jimenez02}.  

The best fit values of the model parameters from OHD are determined by minimizing the quantity 
\begin{equation}
\chi^2_{OHD}= \sum^{18}_{j=1} \frac{\left[ H_{th}(z_j,\mathbf{\theta})-H_{obs}(z_j)\right]^2}{\sigma^2_{H_{obs}}(z_j)},
\label{Eq:COHD}
\end{equation}
where $\sigma^2_H$ are the measurement variances, and $\mathbf{\theta}$ corresponds to the free parameters of the cosmological model. 

\subsection{Long Gamma-Ray Bursts (LGRBs)}

In addition, we use 9 LGRBs  with redshift in the range $1.547 \leq z \leq  3.57$ recently calibrated in Ref. \cite{Yonetoku12} through the Type I Fundamental Plane defined by the correlation between the spectral peak energy $E_p$, the peak luminosity $L_p$, and the luminosity time $T_L\equiv E_{iso}/L_p$, where $E_{iso}$ is the isotropic energy. This calibration is one of several proposals to calibrate GRBs in an cosmology- independent way, required to use them in cosmological tasks. Here, we want to point out that the election of this sample is based on the fact that this compilation leads to stronger constraints due to the control of systematic errors. See Ref. \cite{Yonetoku12} for further details about the calibration. To know more about the state of the art regarding the calibrations performed in an cosmology-independent way see for example Refs. \cite{Kodama08,Liang08,Wei09,Wei10,Wang08,Cardone09}; to go deeper into the debate about the use of GRBs for cosmological purposes, see Refs. \cite{Cuesta,Liang10,Freitas11,Graziani,Collazzi,Butler,Sha,Butler10}.

The $\chi^2$ function for the GRBs data is defined similarly to the SNe Ia data as
\begin{equation}
\chi^2_{LGRBs}= \mathrm{\Delta} \mathrm{F} \cdot \mathrm{C^{-1}} \cdot \mathrm{\Delta} \mathrm{F},
\label{Eq:CLGRB}
\end{equation}
where $\mathrm{F}$ corresponds to the distance modulus given by the Eq. (\ref{Eq:muSN}). As in the case of the SNe Ia sample, we marginalize the $\chi^2_{LGRBs}$ over $\mu_0$.

\subsection{Statistical Method}

To estimate the cosmological parameters of the nonlinear magnetic scenarios, we use a Markov Chain Monte Carlo (MCMC) code. The MCMC method is an algorithm extensively used to sample the parameter space that allows to obtain narrower constraints on the model parameters with the only complication of approaching correctly the convergence of the chain. In particular, our code addresses this issue following the prescription developed and fully described in \cite{Dunkley05}. For a further description on MCMC methods see \cite{Berg,MacKay,Neal} and references therein.

The method is fairly standard. By using our MCMC code, we minimize the $\chi^2$ function thus obtaining the best fit of model parameters from  observational data. This minimization is equivalent to maximize the likelihood function $\mathcal{L}(\theta) \propto \exp [-\chi^2(\theta)/2]$ where $\theta$ is the vector of model parameters. For the nonlinear magnetic scenarios, $\theta$ corresponds to $\Omega_m$ and $h$ for the case when we use the observational Hubble data (OHD) and when we use the combination of all observational data sets, otherwise, $\theta$ corresponds to $\Omega_m$. The expression for $\chi^2(\theta)$ depends on the dataset used, see Eqs. (\ref{Eq:CSNIa}), (\ref{Eq:COHD}) and (\ref{Eq:CLGRB}).


On the other hand, in order to study the influence of a prior on $\Omega_m$, we shall analyze two main cases. In the first one, no prior will be assumed, while in the second we include
a Gaussian prior on $\Omega_m$ from the Planck results, $\Omega_m=0.315\pm 0.017$ \cite{PlanckXVI}. Additionally, when we use observational Hubble data we assume a prior on $H_0=73.8\pm 2.4$ from \cite{Riess} and for running our MCMCs we adopt the physical controls $0<\Omega_m<1$ and $0<h<1$.

\section{Results and Discussion}

\begin{table*}
\caption{Summary of the best estimates of model parameters for the scenario with $\alpha=-1$. We present the best estimates obtained by assuming a prior on $\Omega_m$ from the first Planck results \cite{PlanckXVI} as well as the ones obtained without assuming any prior on $\Omega_m$. The errors are at $68.3\%$ confidence level.}
\label{Table:1}
\centering
\ra{1.5}
\begin{tabular}{@{}lccccccccc@{}}\hline
& \multicolumn{3}{c}{\textbf{With prior on $\Omega_m$}} & \hphantom &\multicolumn{3}{c}{\textbf{Without prior on $\Omega_m$}} &
 \\ \hline
& $\Omega_m$ & $h$ & $\chi^2$ && $\Omega_m$ & $h$ & $\chi^2$\\ \hline
\textbf{OHD} & $0.322^{+0.023}_{-0.023}$& $0.766^{+0.026}_{-0.025}$ & 19.848 && $0.349^{+0.054}_{-0.052}$ &$0.752^{+0.035}_{-0.032}$ & 19.148 \\
\textbf{SNe Ia} & $0.363^{+0.021}_{-0.020}$& $--$ & 587.419&& $0.488^{+0.051}_{-0.049}$& $--$ &561.269\\
\textbf{LGRBs }& $0.314^{+0.026}_{-0.026}$& $--$ & 11.206 && $(0.010,1.0)$& $--$&10.547\\ 
\textbf{Combination} & $0.361^{+0.020}_{-0.019}$& $0.747^{+0.025}_{-0.024}$& 613.444&& $0.438^{+0.038}_{-0.037}$& $0.714^{+0.027}_{-0.026}$& 595.089\\ \hline
\end{tabular}
\end{table*}

\begin{table*}
\caption{Summary of the best estimates of model parameters for the scenario with $\alpha=-1/4$. We present the best estimates obtained by assuming a prior on $\Omega_m$ from the first Planck results \cite{PlanckXVI} as well as the ones obtained without assuming any prior on $\Omega_m$. The errors are at $68.3\%$ confidence level.}
\label{Table:2}
\centering
\ra{1.5}
\begin{tabular}{@{}lccccccccc@{}}\hline
& \multicolumn{3}{c}{\textbf{With prior on $\Omega_m$}} & \hphantom & \multicolumn{3}{c}{\textbf{Without prior on $\Omega_m$}} &
\\ \hline
& $\Omega_m$ & $h$ & $\chi^2_r$ && $\Omega_m$ & $h$ & $\chi^2_r$\\ \hline
\textbf{OHD} & $0.314^{+0.024}_{-0.023}$& $0.731^{+0.024}_{-0.024}$ & 15.034 && $0.309^{+0.056}_{-0.055}$ &$0.732^{+0.032}_{-0.029}$ & 14.984 \\
\textbf{SNe Ia} & $0.327^{+0.023}_{-0.023}$& $--$ & 556.515&& $0.380^{+0.057}_{-0.056}$& $--$ &553.955\\
\textbf{LGRBs }& $0.315^{+0.026}_{-0.026}$& $--$ & 10.900 && $(0.158,0.951)$& $--$&10.549\\ 
\textbf{Combination} & $0.324^{+0.022}_{-0.021}$& $0.726^{+0.024}_{-0.023}$& 578.099&& $0.346^{+0.040}_{-0.040}$& $0.718^{+0.027}_{-0.025}$& 577.169\\ \hline
\end{tabular}
\end{table*}

\begin{table*}
\caption{Summary of the best estimates of model parameters for the scenario with $\alpha=-1/8$. We present the best estimates obtained by assuming a prior on $\Omega_m$ from the first Planck results \cite{PlanckXVI} as well as the ones obtained without assuming any prior on $\Omega_m$. The errors are at $68.3\%$ confidence level.}
\label{Table:3}
\centering
\ra{1.5}
\begin{tabular}{@{}lccccccccc@{}}\hline
& \multicolumn{3}{c}{\textbf{With prior on $\Omega_m$}} &\hphantom &  \multicolumn{3}{c}{\textbf{Without prior on $\Omega_m$}} &
 \\ \hline
& $\Omega_m$ & $h$ & $\chi^2_r$ && $\Omega_m$ & $h$ & $\chi^2_r$\\ \hline
\textbf{OHD} & $0.311^{+0.024}_{-0.024}$& $0.721^{+0.023}_{-0.024}$ & 15.667 && $0.293^{+0.059}_{-0.055}$ &$0.726^{+0.031}_{-0.028}$ & 15.361 \\
\textbf{SNe Ia} & $0.320^{+0.023}_{-0.023}$& $--$ & 553.865&& $0.345^{+0.060}_{-0.058}$& $--$ &553.427\\
\textbf{LGRBs }& $0.315^{+0.026}_{-0.026}$& $--$ & 10.840 && $(0.029, 1.0)$& $--$&10.518\\ 
\textbf{Combination} & $0.316^{+0.022}_{-0.021}$& $0.719^{+0.023}_{-0.023}$& 575.919&& $0.318^{+0.043}_{-0.039}$& $0.718^{+0.027}_{-0.025}$&575.912 \\ \hline
\end{tabular}
\end{table*}

\begin{figure*}
  \centering
  \includegraphics[width=0.47\textwidth]{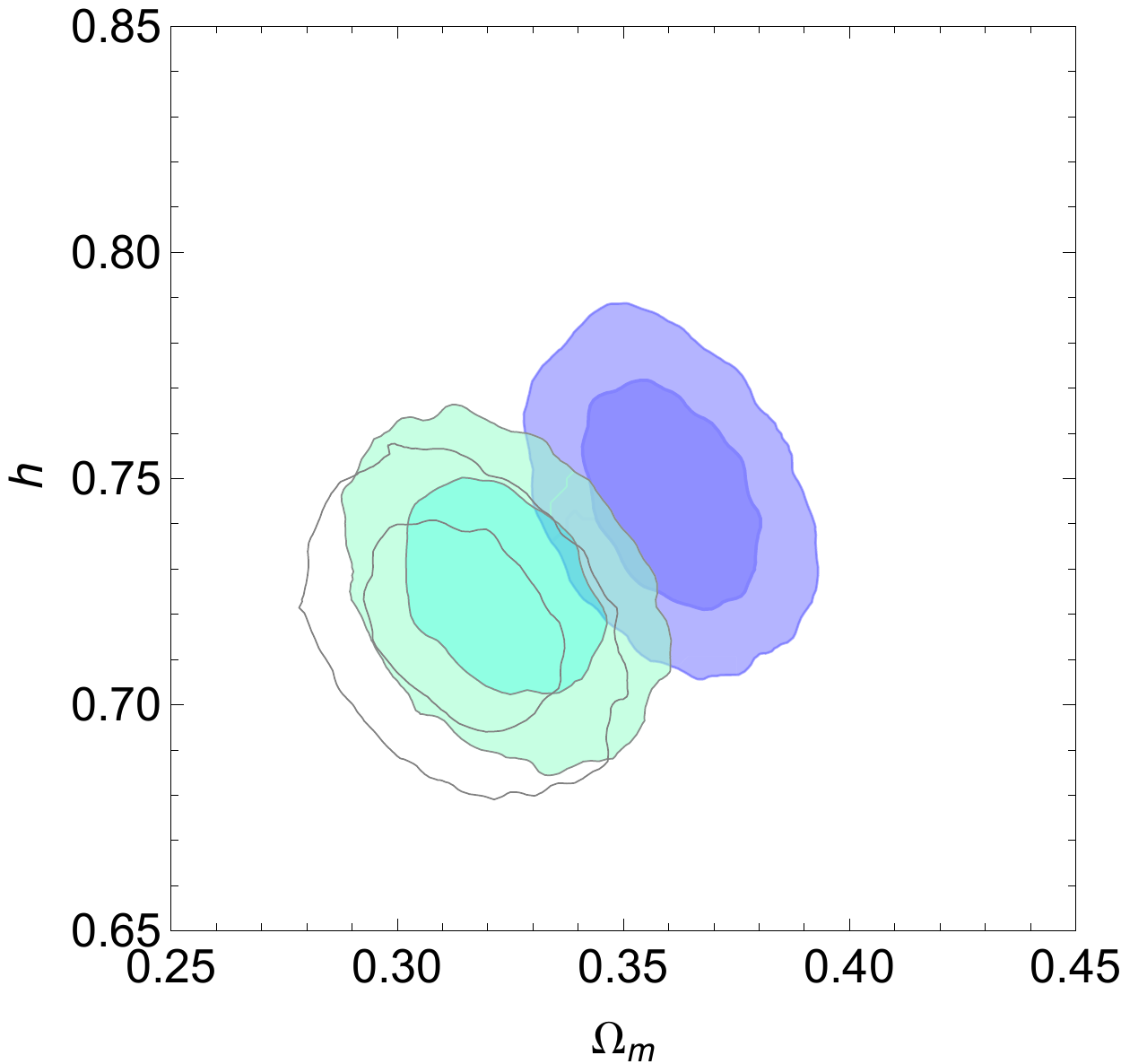}
  \hskip .8cm
    \includegraphics[width=0.45\textwidth]{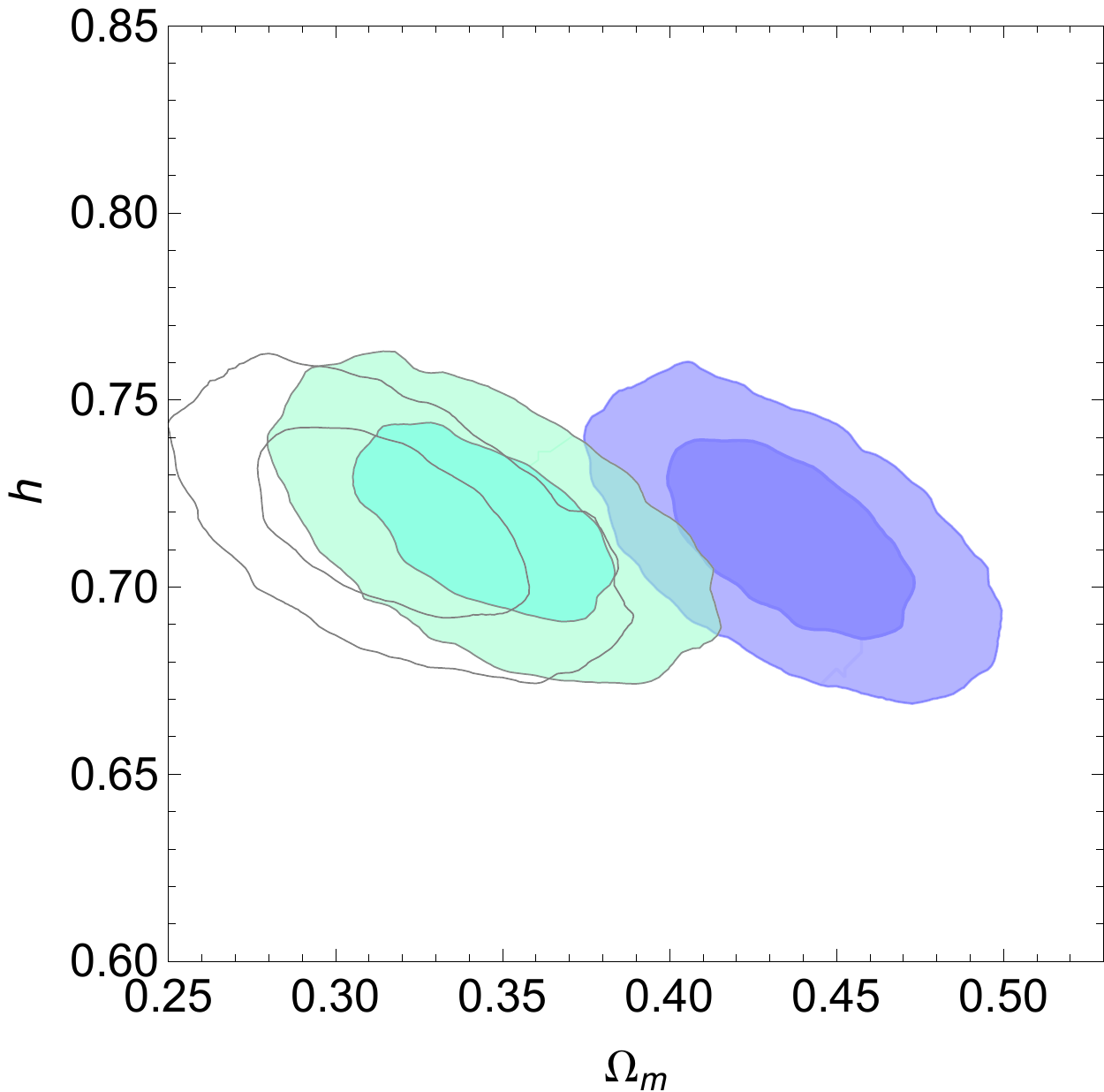}
  \caption{1$\sigma$ and 2$\sigma$ contours in the $\Omega_m$-$h$ parameter space coming from the combination of all observational data. (\textit{Left Panel.}) These confidence regions have been obtained considering a prior on $\Omega_m$ from the Planck results. The blue contours correspond to the nonlinear magnetic universe with $\alpha=-1$; the green contours correspond to the scenario with $\alpha=-1/4$; the contours in solid line corresponds to the scenario with $\alpha=-1/8$.  (\textit{Right Panel.}) In this case, the previous color code also holds but, the contours are obtained without assuming any prior on $\Omega_m$.}
  \label{Fig:WithoutPriorWithPrior}
\end{figure*}

\begin{figure}
  \centering
  \includegraphics[width=0.5\textwidth]{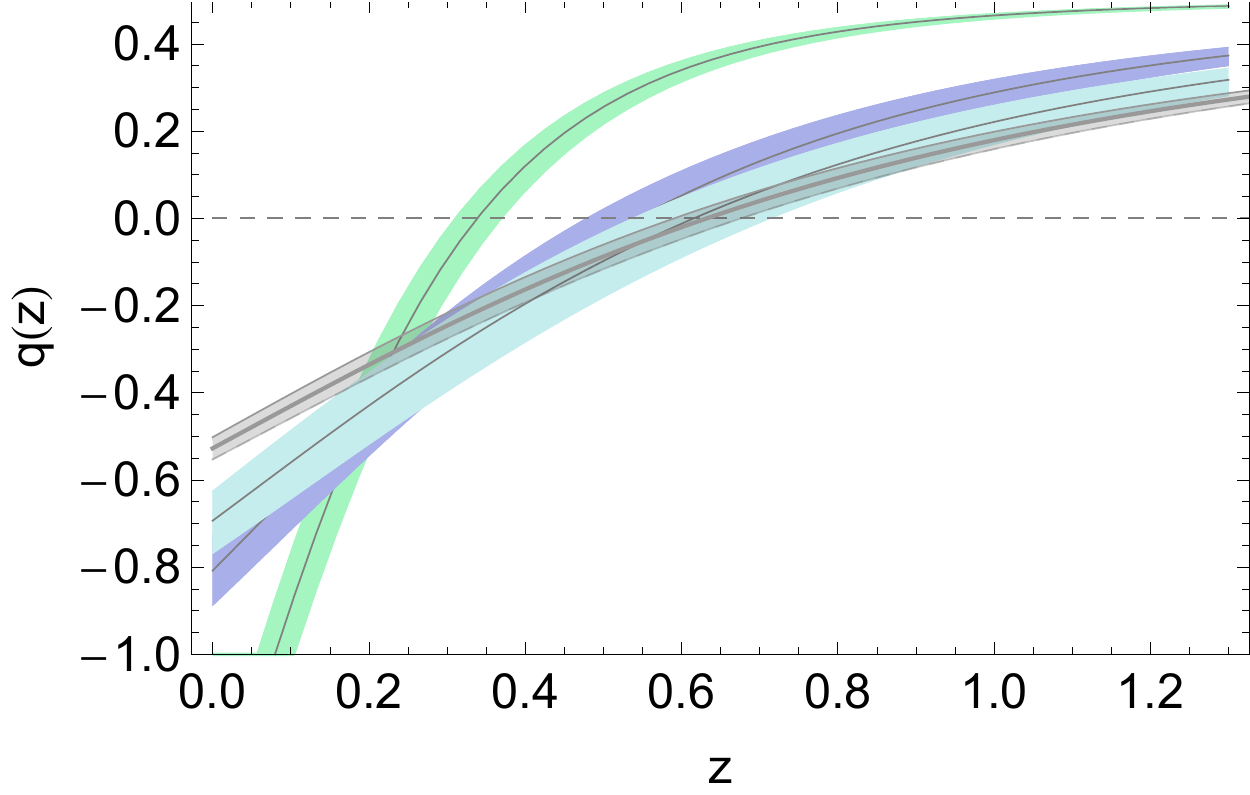}
  \caption{Deceleration parameter $q(z)$ evolving with redshift for the nonlinear magnetic scenarios with $\alpha=-1$ (shaded tight region in mint color), $\alpha=-1/4$ (shaded region in violet) and $\alpha=-1/8$ (shaded region in blue) along with the 1$\sigma$ errors from results from the combination of all observational data sets obtained without any prior and, for comparison, the respective deceleration parameter (shaded region in gray) from $\Lambda$CDM model by using $\Omega_m$ from the Planck results \cite{PlanckXVI}.}
  \label{Fig:q}
\end{figure}


The best fits for the parameters $\Omega_m$ and $h$ for the nonlinear magnetic scenarios with $\alpha=-1$, $\alpha=-1/4$ and $\alpha=-1/8$, as well as the corresponding $\chi^2$, are shown in Tables \ref{Table:1}, \ref{Table:2} and \ref{Table:3}, respectively. 
 
Table \ref{Table:1} contains  the best fits for $\Omega_m$ and $h$ for the scenario with $\alpha=-1$ obtained from OHD, SNe Ia, LGRBs and the Combination of all data sets by assuming a Gaussian prior on $\Omega_m$ and also, without assuming any prior on $\Omega_m$. Table \ref{Table:2} also contains the same fits but now for the scenario with $\alpha=-1/4$ and the Table \ref{Table:3} contains the best fits for the scenario with $\alpha=-1/8$. 

In Table \ref{Table:1}, \ref{Table:2} and \ref{Table:3} can be observed immediately that $\Omega_m$ is poorly constrained by LGRBs,  specially when any prior on $\Omega_m$ is assumed. For the scenario with $\alpha=-1$, $\Omega_m$ is restricted to the interval $(0.010, 1.0)$, for the one with $\alpha=-1/4$, to the interval $(0.158, 0.951)$ and finally, for the scenario with $\alpha=-1/8$ to the interval $(0.029,1.0)$. However, when we assumed a prior on $\Omega_m$, good constraints for all the scenarios were obtained. Notice also that in this last case, the presence of a prior on $\Omega_m$ pushes the nonlinear electromagnetic matter to contribute the total matter content allowing at the same time, a better agreement of $\Omega_m$ with the reported value $\Omega_m=0.315\pm 0.017$ by \cite{PlanckXVI}. 
The corresponding 68 $\%$ and 95 $\%$ likelihood contours from the adjustments by using the combination of all observational data sets are shown in the $\Omega_m$-$h$ parameter space in the Figure \ref{Fig:WithoutPriorWithPrior}. 

As can be noted from Tables \ref{Table:1}, \ref{Table:2} and \ref{Table:3},  SNe Ia data as well as the combination of all data sets yield tighter confidence regions, which is reflected in smaller errors in the best fits. 

On the other hand, if the nonlinear magnetic matter, $\Omega_B$, were sufficient to drive the cosmic acceleration, it would be expected that its contribution to the total matter content were significant, around  68$\%$, as Planck results suggest \cite{PlanckXVI}. In order to estimate such contribution, we use the normalization condition, Eq. (\ref{Eq:norm}), and the best fits for $\Omega_m$. From the combination of all observational data sets, not assuming any prior, $\Omega_B$ is in the interval $0.524 \leq \Omega_B \leq 0.599$ for the scenario with $\alpha=-1$, in the interval $0.614 \leq \Omega_B \leq 0.694$ for the scenario with $\alpha=-1/4$ and in the interval $0.640 \leq \Omega_B \leq 0.722$ for the scenario with $\alpha=-1/8$; the two latest are in better agreement with the recent results of Planck. Considering $\Lambda$CDM model as the most accepted one, the fact that nonlinear electromagnetic matter approaches it via an appropiate value for $\alpha$, might be an indication of which the origin of $\Lambda$ is. The analysis for the case $\alpha=-1/8$, whose results are shown in Table \ref{Table:3}, confirms that a smaller $\alpha$ renders a better fit. Note that these results, approach much more the results from the $\Lambda$CDM scenario than, for example, the ones from the scenario with $\alpha=-1$.

Regarding the deceleration parameter $q(z)$, the values obtained at $z=0$ using the best fits from each observational set, are presented in Table \ref{Table:4}. The evolution of the deceleration parameter $q$ with $z$ for the scenarios with $\alpha=-1$, $\alpha=-1/4$, and $\alpha=-1/8$ obtained from the combination of all observational data sets, without assuming any prior on $\Omega_m$, can be seen in Figure \ref{Fig:q} as well as the deceleration parameter for the $\Lambda$CDM model assuming $\Omega_m$ from \cite{PlanckXVI}.
\begin{table*}
\caption{Summary of the best estimates of deceleration parameter $q_0$ from the combination of all observational data sets with and without a prior on $\Omega_m$ (prior from the first Planck results \cite{PlanckXVI}). The errors are at $68.3\%$ confidence level.}
\label{Table:4}
\centering
\ra{1.5}
\begin{tabular}{@{}lccccc@{}}\hline
 &\textbf{With prior on $\Omega_m$} & \hphantom & \textbf{Without prior on $\Omega_m$} 
 \\ \hline
Model & $q_0$   && $q_0$  \\ \hline
$\alpha=-1$ & $-1.736^{+0.068}_{-0.066}$ && $-1.468^{+0.131}_{-0.130}$   \\
$\alpha=-1/4$ & $-0.852^{+0.043}_{-0.042}$  && $-0.808^{+0.080}_{-0.079}$ \\
$\alpha=-1/8$& $-0.698^{+0.039}_{-0.038}$  && $-0.694^{+0.075}_{-0.069}$ \\ \hline
\end{tabular}
\end{table*}
Note from these figures that the nonlinear magnetic scenarios with $\alpha=-1/4$ and $\alpha=-1/8$ reproduce well the trend of an accelerated expansion scenario driven by a cosmological constant with a transition occurring around $z=0.5$.
 
Finally, using the estimations for $\Omega_B$,  we are able to evaluate the current NLED coupling constant $\gamma$ using Eq. (\ref{Eq:rhoB}). In \cite{Novello2} the authors assumed
that the DE density is $\Omega_{\text{de}}\approx 0.7$ and made an estimation of  $- \gamma=\hbar \mu^4\approx 3.74 \times 10^{-28} \text{g/cm}^3$.

We get an estimation of $\hbar \mu^4$ using our result for $\Omega_B$ and considering  that $B_0$ is attached to the cosmic microwave background (CMB) radiation, 

\begin{equation}
\Omega_{\text{rad}}= \frac{\rho_{\text{rad}}}{\rho_{c,0}}= \frac{B_0^2}{2 \rho_{c,0}}. 
\end{equation}

The resulting coupling constant $\gamma$ from Eq. (\ref{Eq:rhoB}) amounts to
\begin{equation}
- \gamma = \frac{\Omega_B \rho_{c,0}}{(2^2\Omega_{\rm {rad}} \rho_{c,0})^{\alpha}}.
\label{gamma}
\end{equation} 

We will take the value of the radiation density $\Omega_{rad}=2.47 \times 10^{-5} h^{-2}$. In the case $\alpha =-1$, we parametrize $\gamma$ following \cite{Novello2} as $\gamma=\hbar^2 \mu^8$.
Taking the value obtained from the combination of all observational data sets obtained without a prior on $\Omega_m$, $\Omega_B=0.562^{+0.037}_{-0.038}$ and $h=0.714^{+0.027}_{-0.026}$, we obtain
$-\gamma=(1.089^{+0.109}_{-0.108})  \times 10^{-4} \rho_{c,0}^2$ as the coupling constant and  $\hbar \mu^4= (1.044^{+0.052}_{-0.052}) \times 10^{-2} \rho_{c,0}$, one hundredth times smaller than the critical density.

In the case $\alpha=-1/4$, substituting in Eq. (\ref{gamma}),  
from the combination of all observational data sets obtained without a prior on $\Omega_m$, $\Omega_B=0.654^{+0.040}_{-0.040}$ and $h=0.718^{+0.027}_{-0.025}$,
the result for the coupling constant is $- \gamma=(0.077^{+0.002}_{-0.001}) \rho_{c,0}^{5/4} $ or in energy density units  $\hbar \mu^4= (1.286^{+0.066}_{-0.065}) \times 10^{-1} \rho_{c,0}$, one order of magnitude larger than the one with $\alpha =-1$.
As it is mentioned in \cite{Medeiros2012}, it is still difficult to achieve measurements with that precision at present. 

\section{Conclusions}
  
As a phenomenological approach to describe DE, it is interesting to study nonlinear magnetic scenarios with a
Lagrangian of the form $\Ln= \gamma F^{\alpha}$. We performed the adjustment of $\Omega_m$ parameter with three probes: SNe Ia, LGRBs and the Hubble parameter measurements. Technical difficulties lead us to consider the parameter $\alpha$ fixed instead of depending on redshift, and it turned out that  $\alpha=-1/4$  and $\alpha=-1/8$ reproduce pretty well the current observational data. 
 
The best fit for the magnetic component obtained from the combination of all observational data sets is $\Omega_B=0.562^{+0.037}_{-0.038}$ for the scenario in which $\alpha=-1$,  $\Omega_B=0.654^{+0.040}_{-0.040}$ for the one with $\alpha=-1/4$ and $\Omega_B=0.683^{+0.039}_{-0.043}$ for the one with $\alpha=-1/8$
These results allow us to conclude that the nonlinear magnetic matter  could play the role of DE. 
  
In general, the adjustments of $\Omega_m$ and $h$ for the scenario with $\alpha=-1/4$ and for the one with $\alpha=-1/8$ are considerably better than the one with $\alpha=-1$. In addition, although Eq. (\ref{Eq.5}) sets an upper bound for the value of $\alpha$ in order to  produce accelerated expansion, from Eq. (\ref{Eq:alpha}) and our fits from the combination of all observational data sets for the scenario with $\alpha=-1/4$ without assuming any prior on $\Omega_m$, we obtain a bound for $\alpha < 0.368$. A similar bound  of $\alpha<0.385$ can be calculated from the $\Lambda$CDM model.

In spite that we obtained poor constraints for the $\Omega_m$ parameter from LGRBs data without assuming any prior, we should keep in mind that the use of GRBs as cosmological probes is still in debate and LGRBs data are not as reliable as SNe Ia and OHD; however they can give a general idea of the evolution and behaviour of cosmological models at high redshifts. 

On the other hand, regarding LGRBs, notice from the value of $\chi^2$ in Table \ref{Table:1} ($\alpha=-1$) that  $\Omega_m $ is better adjusted than in Table \ref{Table:3} ($\alpha=-1/8$). Remember that a good adjustment is such that $\chi^2$ is closest to the number of data in the sample. The opposite occurs with SNe Ia:  $\Omega_m $ is better adjusted for $\alpha=-1/8$ (see Table \ref{Table:3}) 
than for $\alpha=-1$ (Table \ref{Table:1} ).
If we relate this result with the different redshift ranges that correspond to these probes, $1.547 < z < 3.57$ for LGRBs
and $0.015 < z < 1.4$ for SNe Ia, the difference in the adjustments might indicate that for large redshift the EoS with $\alpha=-1$ models better the cosmic fluid than $\alpha=-1/4$. While for near epochs, a better description is accomplished with $\alpha=-1/4$. 
This result might point to considering the EoS parameter $w(z)$ as redshift dependent. 
 
Finally, although our analysis, that reduces to a perfect fluid one with a constant EoS-parameter, may overlap with some existing in the literature, e.g., with the presented in \cite{Medeiros2012}, in this work we have used the most recent compilation of SNe Ia released by the SCP, unlike the referred work in which it  has been used the Union compilation which only includes 307 data points. Additionally, we have considered direct Hubble parameter measurements and LGRBs data which have extended the range of redshift of study. Besides, we would like to point out that our test was done employing a MCMC method which is more refined one than a standard  $\chi^2$ minimization, thus leading more reliable results.  

\begin{acknowledgements}
A. M. acknowledges financial support from CONACyT (Mexico) through a Ph.D. grant.  N.B. acknowledges partial support by Conacyt, Project 166581. We also acknowledge to the anonymous referee whose suggestions lead to improve our work.
\end{acknowledgements}

\appendix
\section{Scaling between the scale factor $a$ and the electromagnetic invariant $F$ }

The energy conservation $T_{;\mu}^{\mu \nu} = 0$, leads to the equation
\begin{equation}
\dot{\rho} + 3 H (\rho+p)=0,
\end{equation}
which also can be derived from Eq. (\ref{Eq:FriedmannEqs}).  So, using the expressions of $\rho$ and $p$, Eq. (\ref{Eq:EnergyDensity}),  in terms of the electromagnetic Lagrangian, the scaling between the scale factor $a$ and the electromagnetic invariant $F$ can be determined
for a Lagrangian with arbitrary dependence on the two electromagnetic invariants $\Ln(F,G)$ as
\begin{equation}
-\dot{F}\Ln_F + 3 \left({\frac{\dot{a}}{a}}\right) \left({- \frac{4}{3} (2E^2+2B^2)\Ln_F}\right)=0. 
\end{equation}

Now, if one restricts to the case $G=0$ (i.e. no electric field $E=0$), then $F=2B^2$ and
\begin{equation}
-\Ln_F \left\{\dot{F} + 3 \left({\frac{\dot{a}}{a}}\right) \left({ \frac{4}{3}F}\right)\right\}=0,
\end{equation}
whose solution, given by $Fa^4=$const, is independent of the particular form of $\Ln(F)$.

\section{The scale factor as a function of time}

The expressions of Friedmann equations for the nonlinear magnetic terms are
\begin{eqnarray}
\left({\frac{\dot{a}}{a}}\right)^2&=& - \frac{\Ln}{3}, \nonumber\\
\frac{\ddot{a}}{a}&=& - \frac{1}{3} (\Ln-2F\Ln_F).\nonumber\\
\label{FriedmannEqs2}
\end{eqnarray}
Knowing that $\Ln a^{4 \alpha}=$const, and using the Friedmann equations, the expression for $a(t)$ can be determined. Let us consider the following derivative,
\begin{equation}
\frac{d}{dt} \left({a^{(4 \alpha -1)} \dot{a}}\right)= a^{4 \alpha} \left\{{ \frac{\ddot{a}}{a} + (4 \alpha-1) \frac{\dot{a}^2}{a^2}}\right\},
\end{equation}
and substituting Friedmann's equation, Eq. (\ref{FriedmannEqs2}), we realize that the right hand term  is constant,
\begin{equation}
\frac{d}{dt} \left({a^{(4 \alpha -1)} \dot{a}}\right)= {- \frac{2 \alpha \Ln a^{4 \alpha}}{3}} = \text{const}.
\end{equation}

Finally, integrating for $a(t)$, it is obtained that $a(t)= {\rm const}(t-t_0)^{1/2 \alpha}$.

\bibliography{biblio2}

\begin{thebibliography}{53}
\expandafter\ifx\csname natexlab\endcsname\relax\def\natexlab#1{#1}\fi
\expandafter\ifx\csname bibnamefont\endcsname\relax
  \def\bibnamefont#1{#1}\fi
\expandafter\ifx\csname bibfnamefont\endcsname\relax
  \def\bibfnamefont#1{#1}\fi
\expandafter\ifx\csname citenamefont\endcsname\relax
  \def\citenamefont#1{#1}\fi
\expandafter\ifx\csname url\endcsname\relax
  \def\url#1{\texttt{#1}}\fi
\expandafter\ifx\csname urlprefix\endcsname\relax\def\urlprefix{URL }\fi
\providecommand{\bibinfo}[2]{#2}
\providecommand{\eprint}[2][]{\url{#2}}

\bibitem[{\citenamefont{Novello et~al.}(2004)\citenamefont{Novello,
  Perez~Bergliaffa, and Salim}}]{Novello2}
\bibinfo{author}{\bibfnamefont{M.}~\bibnamefont{Novello}},
  \bibinfo{author}{\bibfnamefont{S.~E.} \bibnamefont{Perez~Bergliaffa}},
  \bibnamefont{and} \bibinfo{author}{\bibfnamefont{J.}~\bibnamefont{Salim}},
  \bibinfo{journal}{Phys.Rev.} \textbf{\bibinfo{volume}{D69}},
  \bibinfo{pages}{127301} (\bibinfo{year}{2004}), \eprint{astro-ph/0312093}.

\bibitem[{\citenamefont{Novello et~al.}(2007)\citenamefont{Novello, Goulart,
  Salim, and Perez~Bergliaffa}}]{Novello3}
\bibinfo{author}{\bibfnamefont{M.}~\bibnamefont{Novello}},
  \bibinfo{author}{\bibfnamefont{E.}~\bibnamefont{Goulart}},
  \bibinfo{author}{\bibfnamefont{J.}~\bibnamefont{Salim}}, \bibnamefont{and}
  \bibinfo{author}{\bibfnamefont{S.}~\bibnamefont{Perez~Bergliaffa}},
  \bibinfo{journal}{Class.Quant.Grav.} \textbf{\bibinfo{volume}{24}},
  \bibinfo{pages}{3021} (\bibinfo{year}{2007}), \eprint{gr-qc/0610043}.

\bibitem[{\citenamefont{Vollick}(2008)}]{Vollick2008}
\bibinfo{author}{\bibfnamefont{D.~N.} \bibnamefont{Vollick}},
  \bibinfo{journal}{Phys.Rev.} \textbf{\bibinfo{volume}{D78}},
  \bibinfo{pages}{063524} (\bibinfo{year}{2008}), \eprint{0807.0448}.

\bibitem[{\citenamefont{Labun and Rafelski}(2010)}]{Labun2010}
\bibinfo{author}{\bibfnamefont{L.}~\bibnamefont{Labun}} \bibnamefont{and}
  \bibinfo{author}{\bibfnamefont{J.}~\bibnamefont{Rafelski}},
  \bibinfo{journal}{Phys.Rev.} \textbf{\bibinfo{volume}{D81}},
  \bibinfo{pages}{065026} (\bibinfo{year}{2010}), \eprint{0811.4467}.

\bibitem[{\citenamefont{Dyadichev et~al.}(2002)\citenamefont{Dyadichev,
  Gal'tsov, Zorin, and Zotov}}]{Dyadichev2002}
\bibinfo{author}{\bibfnamefont{V.}~\bibnamefont{Dyadichev}},
  \bibinfo{author}{\bibfnamefont{D.}~\bibnamefont{Gal'tsov}},
  \bibinfo{author}{\bibfnamefont{A.}~\bibnamefont{Zorin}}, \bibnamefont{and}
  \bibinfo{author}{\bibfnamefont{M.~Y.} \bibnamefont{Zotov}},
  \bibinfo{journal}{Phys.Rev.} \textbf{\bibinfo{volume}{D65}},
  \bibinfo{pages}{084007} (\bibinfo{year}{2002}), \eprint{hep-th/0111099}.

\bibitem[{\citenamefont{Elizalde et~al.}(2003)\citenamefont{Elizalde, Lidsey,
  Nojiri, and Odintsov}}]{Elizalde2003}
\bibinfo{author}{\bibfnamefont{E.}~\bibnamefont{Elizalde}},
  \bibinfo{author}{\bibfnamefont{J.~E.} \bibnamefont{Lidsey}},
  \bibinfo{author}{\bibfnamefont{S.}~\bibnamefont{Nojiri}}, \bibnamefont{and}
  \bibinfo{author}{\bibfnamefont{S.~D.} \bibnamefont{Odintsov}},
  \bibinfo{journal}{Phys.Lett.} \textbf{\bibinfo{volume}{B574}},
  \bibinfo{pages}{1} (\bibinfo{year}{2003}), \eprint{hep-th/0307177}.

\bibitem[{\citenamefont{Beltran~Jimenez and Maroto}(2008)}]{Jimenez:2008au}
\bibinfo{author}{\bibfnamefont{J.}~\bibnamefont{Beltran~Jimenez}}
  \bibnamefont{and} \bibinfo{author}{\bibfnamefont{A.~L.}
  \bibnamefont{Maroto}}, \bibinfo{journal}{Phys.Rev.}
  \textbf{\bibinfo{volume}{D78}}, \bibinfo{pages}{063005}
  (\bibinfo{year}{2008}), \eprint{0801.1486}.

\bibitem[{\citenamefont{Beltran~Jimenez and
  Maroto}(2009{\natexlab{a}})}]{Jimenez:2008nm}
\bibinfo{author}{\bibfnamefont{J.}~\bibnamefont{Beltran~Jimenez}}
  \bibnamefont{and} \bibinfo{author}{\bibfnamefont{A.~L.}
  \bibnamefont{Maroto}}, \bibinfo{journal}{JCAP}
  \textbf{\bibinfo{volume}{0903}}, \bibinfo{pages}{016}
  (\bibinfo{year}{2009}{\natexlab{a}}), \eprint{0811.0566}.

\bibitem[{\citenamefont{Beltran~Jimenez and
  Maroto}(2009{\natexlab{b}})}]{Jimenez:2008er}
\bibinfo{author}{\bibfnamefont{J.}~\bibnamefont{Beltran~Jimenez}}
  \bibnamefont{and} \bibinfo{author}{\bibfnamefont{A.~L.}
  \bibnamefont{Maroto}}, \bibinfo{journal}{AIP Conf.Proc.}
  \textbf{\bibinfo{volume}{1122}}, \bibinfo{pages}{107}
  (\bibinfo{year}{2009}{\natexlab{b}}), \eprint{0812.1970}.

\bibitem[{\citenamefont{Medeiros}(2012)}]{Medeiros2012}
\bibinfo{author}{\bibfnamefont{L.}~\bibnamefont{Medeiros}},
  \bibinfo{journal}{Int.J.Mod.Phys.} \textbf{\bibinfo{volume}{D23}},
  \bibinfo{pages}{1250073} (\bibinfo{year}{2012}), \eprint{1209.1124}.

\bibitem[{\citenamefont{Pleba{\'n}ski}(1970)}]{Pleban}
\bibinfo{author}{\bibfnamefont{J.}~\bibnamefont{Pleba{\'n}ski}},
  \emph{\bibinfo{title}{Lectures on non-linear electrodynamics: an extended
  version of lectures given at the Niels Bohr Institute and NORDITA,
  Copenhagen, in October 1968}} (\bibinfo{publisher}{NORDITA},
  \bibinfo{year}{1970}).

\bibitem[{\citenamefont{Mottola}()}]{Mottola}
\bibinfo{author}{\bibfnamefont{E.}~\bibnamefont{Mottola}},
  \bibinfo{journal}{Proceedings of the XLVth Rencontres de Moriond,2010
  Cosmology, edited by E. Auge, J. Dumarchez and J. Tran Tranh an, The Gioi
  Publishers, Vietnam (2010)}  (????), \eprint{1103.1613}.

\bibitem[{\citenamefont{Mosquera~Cuesta
  et~al.}(2007)\citenamefont{Mosquera~Cuesta, Salim, and Novello}}]{Mosquera2}
\bibinfo{author}{\bibfnamefont{H.~J.} \bibnamefont{Mosquera~Cuesta}},
  \bibinfo{author}{\bibfnamefont{J.~M.} \bibnamefont{Salim}}, \bibnamefont{and}
  \bibinfo{author}{\bibfnamefont{M.}~\bibnamefont{Novello}}
  (\bibinfo{year}{2007}), \eprint{astro-ph/0710.5188}.

\bibitem[{\citenamefont{Born and Infeld}(1934)}]{BI}
\bibinfo{author}{\bibfnamefont{M.}~\bibnamefont{Born}} \bibnamefont{and}
  \bibinfo{author}{\bibfnamefont{L.}~\bibnamefont{Infeld}},
  \bibinfo{journal}{Proc.Roy.Soc.Lond.} \textbf{\bibinfo{volume}{A144}},
  \bibinfo{pages}{425} (\bibinfo{year}{1934}).

\bibitem[{\citenamefont{Tolman and Ehrenfest}(1930)}]{Tolman}
\bibinfo{author}{\bibfnamefont{R.~C.} \bibnamefont{Tolman}} \bibnamefont{and}
  \bibinfo{author}{\bibfnamefont{P.}~\bibnamefont{Ehrenfest}},
  \bibinfo{journal}{Phys. Rev.} \textbf{\bibinfo{volume}{36}},
  \bibinfo{pages}{1791} (\bibinfo{year}{1930}).

\bibitem[{\citenamefont{Armendariz-Picon}(2004)}]{ArmendarizP}
\bibinfo{author}{\bibfnamefont{C.}~\bibnamefont{Armendariz-Picon}},
  \bibinfo{journal}{JCAP} \textbf{\bibinfo{volume}{0407}}, \bibinfo{pages}{007}
  (\bibinfo{year}{2004}), \eprint{astro-ph/0405267}.

\bibitem[{\citenamefont{Cembranos et~al.}(2012)\citenamefont{Cembranos,
  Hallabrin, Maroto, and Jareno}}]{Cembranos:2012kk}
\bibinfo{author}{\bibfnamefont{J.}~\bibnamefont{Cembranos}},
  \bibinfo{author}{\bibfnamefont{C.}~\bibnamefont{Hallabrin}},
  \bibinfo{author}{\bibfnamefont{A.}~\bibnamefont{Maroto}}, \bibnamefont{and}
  \bibinfo{author}{\bibfnamefont{S.~N.} \bibnamefont{Jareno}},
  \bibinfo{journal}{Phys.Rev.} \textbf{\bibinfo{volume}{D86}},
  \bibinfo{pages}{021301} (\bibinfo{year}{2012}), \eprint{1203.6221}.

\bibitem[{\citenamefont{Novello}(2005)}]{Novello:2005bj}
\bibinfo{author}{\bibfnamefont{M.}~\bibnamefont{Novello}},
  \bibinfo{journal}{Int.J.Mod.Phys.} \textbf{\bibinfo{volume}{A20}},
  \bibinfo{pages}{2421} (\bibinfo{year}{2005}).

\bibitem[{\citenamefont{Lemoine and Lemoine}(1995)}]{Lemoine:1995vj}
\bibinfo{author}{\bibfnamefont{D.}~\bibnamefont{Lemoine}} \bibnamefont{and}
  \bibinfo{author}{\bibfnamefont{M.}~\bibnamefont{Lemoine}},
  \bibinfo{journal}{Phys.Rev.} \textbf{\bibinfo{volume}{D52}},
  \bibinfo{pages}{1955} (\bibinfo{year}{1995}).

\bibitem[{\citenamefont{Novello et~al.}(2009)\citenamefont{Novello, Araujo, and
  Salim}}]{Novello:2008xp}
\bibinfo{author}{\bibfnamefont{M.}~\bibnamefont{Novello}},
  \bibinfo{author}{\bibfnamefont{A.~N.} \bibnamefont{Araujo}},
  \bibnamefont{and} \bibinfo{author}{\bibfnamefont{J.}~\bibnamefont{Salim}},
  \bibinfo{journal}{Int.J.Mod.Phys.} \textbf{\bibinfo{volume}{A24}},
  \bibinfo{pages}{5639} (\bibinfo{year}{2009}), \eprint{0802.1875}.

\bibitem[{\citenamefont{Chayan~Ranjit}(2013)}]{Chayan2013}
\bibinfo{author}{\bibfnamefont{U.~D.} \bibnamefont{Chayan~Ranjit},
  \bibfnamefont{Shuvendu~Chakraborty}}, \bibinfo{journal}{Astrophys. Space Sci}
  \textbf{\bibinfo{volume}{346}}, \bibinfo{pages}{291} (\bibinfo{year}{2013}),
  \eprint{physics.gen-ph/1304.1281}.

\bibitem[{\citenamefont{Heisenberg and Euler}(1936)}]{EulerHeis}
\bibinfo{author}{\bibfnamefont{W.}~\bibnamefont{Heisenberg}} \bibnamefont{and}
  \bibinfo{author}{\bibfnamefont{H.}~\bibnamefont{Euler}}, \bibinfo{journal}{Z.
  Phys.} \textbf{\bibinfo{volume}{38}}, \bibinfo{pages}{714}
  (\bibinfo{year}{1936}).

\bibitem[{\citenamefont{Pagels and Tomboulis}(1978)}]{Pagels:1978dd}
\bibinfo{author}{\bibfnamefont{H.}~\bibnamefont{Pagels}} \bibnamefont{and}
  \bibinfo{author}{\bibfnamefont{E.}~\bibnamefont{Tomboulis}},
  \bibinfo{journal}{Nucl.Phys.} \textbf{\bibinfo{volume}{B143}},
  \bibinfo{pages}{485} (\bibinfo{year}{1978}).

\bibitem[{\citenamefont{Novello et~al.}(2012)\citenamefont{Novello, Salim, and
  Araujo}}]{Novello4}
\bibinfo{author}{\bibfnamefont{M.}~\bibnamefont{Novello}},
  \bibinfo{author}{\bibfnamefont{J.}~\bibnamefont{Salim}}, \bibnamefont{and}
  \bibinfo{author}{\bibfnamefont{A.~N.} \bibnamefont{Araujo}},
  \bibinfo{journal}{Phys.Rev.} \textbf{\bibinfo{volume}{D85}},
  \bibinfo{pages}{023528} (\bibinfo{year}{2012}).

\bibitem[{\citenamefont{Esposito-Farese
  et~al.}(2010)\citenamefont{Esposito-Farese, Pitrou, and
  Uzan}}]{EspositoFarese:2009aj}
\bibinfo{author}{\bibfnamefont{G.}~\bibnamefont{Esposito-Farese}},
  \bibinfo{author}{\bibfnamefont{C.}~\bibnamefont{Pitrou}}, \bibnamefont{and}
  \bibinfo{author}{\bibfnamefont{J.-P.} \bibnamefont{Uzan}},
  \bibinfo{journal}{Phys.Rev.} \textbf{\bibinfo{volume}{D81}},
  \bibinfo{pages}{063519} (\bibinfo{year}{2010}), \eprint{0912.0481}.

\bibitem[{\citenamefont{Golovnev and Klementev}(2014)}]{Golovnev:2013gpa}
\bibinfo{author}{\bibfnamefont{A.}~\bibnamefont{Golovnev}} \bibnamefont{and}
  \bibinfo{author}{\bibfnamefont{A.}~\bibnamefont{Klementev}},
  \bibinfo{journal}{JCAP} \textbf{\bibinfo{volume}{1402}}, \bibinfo{pages}{033}
  (\bibinfo{year}{2014}), \eprint{1311.0601}.

\bibitem[{\citenamefont{Sola and Stefancic}(2005)}]{Sola:2005et}
\bibinfo{author}{\bibfnamefont{J.}~\bibnamefont{Sola}} \bibnamefont{and}
  \bibinfo{author}{\bibfnamefont{H.}~\bibnamefont{Stefancic}},
  \bibinfo{journal}{Phys.Lett.} \textbf{\bibinfo{volume}{B624}},
  \bibinfo{pages}{147} (\bibinfo{year}{2005}), \eprint{astro-ph/0505133}.

\bibitem[{\citenamefont{Caldwell}(2002)}]{Phantom}
\bibinfo{author}{\bibfnamefont{R.}~\bibnamefont{Caldwell}},
  \bibinfo{journal}{Phys.Lett.} \textbf{\bibinfo{volume}{B545}},
  \bibinfo{pages}{23} (\bibinfo{year}{2002}), \eprint{astro-ph/9908168}.

\bibitem[{\citenamefont{Ade et~al.}(2013{\natexlab{a}})}]{PlanckXXVI}
\bibinfo{author}{\bibfnamefont{P.}~\bibnamefont{Ade}} \bibnamefont{et~al.}
  (\bibinfo{collaboration}{Planck Collaboration})
  (\bibinfo{year}{2013}{\natexlab{a}}), \eprint{astro-ph/1303.5086}.

\bibitem[{\citenamefont{Suzuki et~al.}(2012)\citenamefont{Suzuki, Rubin,
  Lidman, Aldering, Amanullah et~al.}}]{Union21}
\bibinfo{author}{\bibfnamefont{N.}~\bibnamefont{Suzuki}},
  \bibinfo{author}{\bibfnamefont{D.}~\bibnamefont{Rubin}},
  \bibinfo{author}{\bibfnamefont{C.}~\bibnamefont{Lidman}},
  \bibinfo{author}{\bibfnamefont{G.}~\bibnamefont{Aldering}},
  \bibinfo{author}{\bibfnamefont{R.}~\bibnamefont{Amanullah}},
  \bibnamefont{et~al.}, \bibinfo{journal}{Astrophys.J.}
  \textbf{\bibinfo{volume}{746}}, \bibinfo{pages}{85} (\bibinfo{year}{2012}),
  \eprint{astro-ph/1105.3470}.

\bibitem[{\citenamefont{Moresco et~al.}(2012)\citenamefont{Moresco, Verde,
  Pozzetti, Jimenez, and Cimatti}}]{Jimenez12}
\bibinfo{author}{\bibfnamefont{M.}~\bibnamefont{Moresco}},
  \bibinfo{author}{\bibfnamefont{L.}~\bibnamefont{Verde}},
  \bibinfo{author}{\bibfnamefont{L.}~\bibnamefont{Pozzetti}},
  \bibinfo{author}{\bibfnamefont{R.}~\bibnamefont{Jimenez}}, \bibnamefont{and}
  \bibinfo{author}{\bibfnamefont{A.}~\bibnamefont{Cimatti}},
  \bibinfo{journal}{JCAP} \textbf{\bibinfo{volume}{1207}}, \bibinfo{pages}{053}
  (\bibinfo{year}{2012}), \eprint{astro-ph/1201.6658}.

\bibitem[{\citenamefont{Jimenez and Loeb}(2002)}]{Jimenez02}
\bibinfo{author}{\bibfnamefont{R.}~\bibnamefont{Jimenez}} \bibnamefont{and}
  \bibinfo{author}{\bibfnamefont{A.}~\bibnamefont{Loeb}},
  \bibinfo{journal}{Astrophys.J.} \textbf{\bibinfo{volume}{573}},
  \bibinfo{pages}{37} (\bibinfo{year}{2002}), \eprint{astro-ph/0106145}.

\bibitem[{\citenamefont{Tsutsui et~al.}(2012)\citenamefont{Tsutsui, Nakamura,
  Yonetoku, Takahashi, and Morihara}}]{Yonetoku12}
\bibinfo{author}{\bibfnamefont{R.}~\bibnamefont{Tsutsui}},
  \bibinfo{author}{\bibfnamefont{T.}~\bibnamefont{Nakamura}},
  \bibinfo{author}{\bibfnamefont{D.}~\bibnamefont{Yonetoku}},
  \bibinfo{author}{\bibfnamefont{K.}~\bibnamefont{Takahashi}},
  \bibnamefont{and} \bibinfo{author}{\bibfnamefont{Y.}~\bibnamefont{Morihara}}
  (\bibinfo{year}{2012}), \eprint{astro-ph/1205.2954}.

\bibitem[{\citenamefont{{Kodama} et~al.}(2008)\citenamefont{{Kodama},
  {Yonetoku}, {Murakami}, {Tanabe}, {Tsutsui}, and {Nakamura}}}]{Kodama08}
\bibinfo{author}{\bibfnamefont{Y.}~\bibnamefont{{Kodama}}},
  \bibinfo{author}{\bibfnamefont{D.}~\bibnamefont{{Yonetoku}}},
  \bibinfo{author}{\bibfnamefont{T.}~\bibnamefont{{Murakami}}},
  \bibinfo{author}{\bibfnamefont{S.}~\bibnamefont{{Tanabe}}},
  \bibinfo{author}{\bibfnamefont{R.}~\bibnamefont{{Tsutsui}}},
  \bibnamefont{and}
  \bibinfo{author}{\bibfnamefont{T.}~\bibnamefont{{Nakamura}}},
  \bibinfo{journal}{Mon. Not. Roy. Astron. Soc.}
  \textbf{\bibinfo{volume}{391}}, \bibinfo{pages}{L1} (\bibinfo{year}{2008}),
  \eprint{astro-ph/0802.3428}.

\bibitem[{\citenamefont{Liang et~al.}(2008)\citenamefont{Liang, Xiao, Liu, and
  Zhang}}]{Liang08}
\bibinfo{author}{\bibfnamefont{N.}~\bibnamefont{Liang}},
  \bibinfo{author}{\bibfnamefont{W.~K.} \bibnamefont{Xiao}},
  \bibinfo{author}{\bibfnamefont{Y.}~\bibnamefont{Liu}}, \bibnamefont{and}
  \bibinfo{author}{\bibfnamefont{S.~N.} \bibnamefont{Zhang}},
  \bibinfo{journal}{Astrophys. J.} \textbf{\bibinfo{volume}{685}},
  \bibinfo{pages}{354} (\bibinfo{year}{2008}).

\bibitem[{\citenamefont{Wei and Zhang}(2009)}]{Wei09}
\bibinfo{author}{\bibfnamefont{H.}~\bibnamefont{Wei}} \bibnamefont{and}
  \bibinfo{author}{\bibfnamefont{S.~N.} \bibnamefont{Zhang}},
  \bibinfo{journal}{Eur.Phys.J.} \textbf{\bibinfo{volume}{C63}},
  \bibinfo{pages}{139} (\bibinfo{year}{2009}), \eprint{astro-ph/0808.2240}.

\bibitem[{\citenamefont{Wei}(2010)}]{Wei10}
\bibinfo{author}{\bibfnamefont{H.}~\bibnamefont{Wei}}, \bibinfo{journal}{JCAP}
  \textbf{\bibinfo{volume}{1008}}, \bibinfo{pages}{020} (\bibinfo{year}{2010}),
  \eprint{astro-ph/1004.4951}.

\bibitem[{\citenamefont{Wang}(2008)}]{Wang08}
\bibinfo{author}{\bibfnamefont{Y.}~\bibnamefont{Wang}},
  \bibinfo{journal}{Phys.Rev.} \textbf{\bibinfo{volume}{D78}},
  \bibinfo{pages}{123532} (\bibinfo{year}{2008}), \eprint{astro-ph/0809.0657}.

\bibitem[{\citenamefont{Cardone et~al.}(2009)\citenamefont{Cardone,
  Capozziello, and Dainotti}}]{Cardone09}
\bibinfo{author}{\bibfnamefont{V.~F.} \bibnamefont{Cardone}},
  \bibinfo{author}{\bibfnamefont{S.}~\bibnamefont{Capozziello}},
  \bibnamefont{and} \bibinfo{author}{\bibfnamefont{M.~G.}
  \bibnamefont{Dainotti}}, \bibinfo{journal}{Mon. Not. Roy. Astron. Soc.}
  \textbf{\bibinfo{volume}{400}}, \bibinfo{pages}{775} (\bibinfo{year}{2009}),
  ISSN \bibinfo{issn}{1365-2966}.

\bibitem[{\citenamefont{Mosquera~Cuesta
  et~al.}(2008)\citenamefont{Mosquera~Cuesta, Dumet~M., and
  Furlanetto}}]{Cuesta}
\bibinfo{author}{\bibfnamefont{H.~J.} \bibnamefont{Mosquera~Cuesta}},
  \bibinfo{author}{\bibfnamefont{H.}~\bibnamefont{Dumet~M.}}, \bibnamefont{and}
  \bibinfo{author}{\bibfnamefont{C.}~\bibnamefont{Furlanetto}},
  \bibinfo{journal}{JCAP} \textbf{\bibinfo{volume}{0807}}, \bibinfo{pages}{004}
  (\bibinfo{year}{2008}), \eprint{astro-ph/0708.1355}.

\bibitem[{\citenamefont{Liang et~al.}(2010)\citenamefont{Liang, Wu, and
  Zhang}}]{Liang10}
\bibinfo{author}{\bibfnamefont{N.}~\bibnamefont{Liang}},
  \bibinfo{author}{\bibfnamefont{P.}~\bibnamefont{Wu}}, \bibnamefont{and}
  \bibinfo{author}{\bibfnamefont{S.~N.} \bibnamefont{Zhang}},
  \bibinfo{journal}{Phys.Rev.} \textbf{\bibinfo{volume}{D81}},
  \bibinfo{pages}{083518} (\bibinfo{year}{2010}), \eprint{astro-ph/0911.5644}.

\bibitem[{\citenamefont{Freitas et~al.}(2011)\citenamefont{Freitas, Goncalves,
  and Velten}}]{Freitas11}
\bibinfo{author}{\bibfnamefont{R.}~\bibnamefont{Freitas}},
  \bibinfo{author}{\bibfnamefont{S.}~\bibnamefont{Goncalves}},
  \bibnamefont{and} \bibinfo{author}{\bibfnamefont{H.}~\bibnamefont{Velten}},
  \bibinfo{journal}{Phys.Lett.} \textbf{\bibinfo{volume}{B703}},
  \bibinfo{pages}{209} (\bibinfo{year}{2011}), \eprint{astro-ph/1004.5585}.

\bibitem[{\citenamefont{Graziani}(2011)}]{Graziani}
\bibinfo{author}{\bibfnamefont{C.}~\bibnamefont{Graziani}},
  \bibinfo{journal}{New Astron.} \textbf{\bibinfo{volume}{16}},
  \bibinfo{pages}{57} (\bibinfo{year}{2011}), \eprint{astro-ph/1002.3434}.

\bibitem[{\citenamefont{Collazzi et~al.}(2012)\citenamefont{Collazzi, Schaefer,
  Goldstein, and Preece}}]{Collazzi}
\bibinfo{author}{\bibfnamefont{A.~C.} \bibnamefont{Collazzi}},
  \bibinfo{author}{\bibfnamefont{B.~E.} \bibnamefont{Schaefer}},
  \bibinfo{author}{\bibfnamefont{A.}~\bibnamefont{Goldstein}},
  \bibnamefont{and} \bibinfo{author}{\bibfnamefont{R.~D.}
  \bibnamefont{Preece}}, \bibinfo{journal}{Astrophys.J.}
  \textbf{\bibinfo{volume}{747}}, \bibinfo{pages}{39} (\bibinfo{year}{2012}),
  \eprint{astro-ph/1112.4347}.

\bibitem[{\citenamefont{Butler et~al.}(2009)\citenamefont{Butler, Kocevski, and
  Bloom}}]{Butler}
\bibinfo{author}{\bibfnamefont{N.~R.} \bibnamefont{Butler}},
  \bibinfo{author}{\bibfnamefont{D.}~\bibnamefont{Kocevski}}, \bibnamefont{and}
  \bibinfo{author}{\bibfnamefont{J.~S.} \bibnamefont{Bloom}},
  \bibinfo{journal}{Astrophys. J.} \textbf{\bibinfo{volume}{694}},
  \bibinfo{pages}{76} (\bibinfo{year}{2009}).

\bibitem[{\citenamefont{Shahmoradi and Nemiroff}(2011)}]{Sha}
\bibinfo{author}{\bibfnamefont{A.}~\bibnamefont{Shahmoradi}} \bibnamefont{and}
  \bibinfo{author}{\bibfnamefont{R.}~\bibnamefont{Nemiroff}},
  \bibinfo{journal}{Mon.Not.Roy.Astron.Soc.} \textbf{\bibinfo{volume}{411}},
  \bibinfo{pages}{1843} (\bibinfo{year}{2011}), \eprint{astro-ph/0904.1464}.

\bibitem[{\citenamefont{Butler et~al.}(2010)\citenamefont{Butler, Bloom, and
  Poznanski}}]{Butler10}
\bibinfo{author}{\bibfnamefont{N.~R.} \bibnamefont{Butler}},
  \bibinfo{author}{\bibfnamefont{J.~S.} \bibnamefont{Bloom}}, \bibnamefont{and}
  \bibinfo{author}{\bibfnamefont{D.}~\bibnamefont{Poznanski}},
  \bibinfo{journal}{Astrophys. J.} \textbf{\bibinfo{volume}{711}},
  \bibinfo{pages}{495} (\bibinfo{year}{2010}), \eprint{astro-ph/0910.3341}.

\bibitem[{\citenamefont{Dunkley et~al.}(2005)\citenamefont{Dunkley, Bucher,
  Ferreira, Moodley, and Skordis}}]{Dunkley05}
\bibinfo{author}{\bibfnamefont{J.}~\bibnamefont{Dunkley}},
  \bibinfo{author}{\bibfnamefont{M.}~\bibnamefont{Bucher}},
  \bibinfo{author}{\bibfnamefont{P.~G.} \bibnamefont{Ferreira}},
  \bibinfo{author}{\bibfnamefont{K.}~\bibnamefont{Moodley}}, \bibnamefont{and}
  \bibinfo{author}{\bibfnamefont{C.}~\bibnamefont{Skordis}},
  \bibinfo{journal}{Mon. Not. Roy. Astron. Soc.}
  \textbf{\bibinfo{volume}{356}}, \bibinfo{pages}{925} (\bibinfo{year}{2005}),
  \eprint{astro-ph/0405462}.

\bibitem[{\citenamefont{Berg}(2004)}]{Berg}
\bibinfo{author}{\bibfnamefont{B.}~\bibnamefont{Berg}},
  \emph{\bibinfo{title}{Markov Chain Monte Carlo Simulations And Their
  Statistical Analysis: With Web-based Fortran Code}}
  (\bibinfo{publisher}{World Scientific Publishing Company, Incorporated},
  \bibinfo{year}{2004}), ISBN \bibinfo{isbn}{9789812389350}.

\bibitem[{\citenamefont{MacKay}(2003)}]{MacKay}
\bibinfo{author}{\bibfnamefont{D.~J.~C.} \bibnamefont{MacKay}},
  \emph{\bibinfo{title}{Information Theory, Inference and Learning Algorithms}}
  (\bibinfo{publisher}{Cambrdige University Press}, \bibinfo{year}{2003}), ISBN
  \bibinfo{isbn}{0521642981}.

\bibitem[{\citenamefont{Neal}(1993)}]{Neal}
\bibinfo{author}{\bibfnamefont{R.~M.} \bibnamefont{Neal}}, \bibinfo{type}{Tech.
  Rep.} \bibinfo{number}{CRG-TR-93-1}, \bibinfo{institution}{Dept. of Computer
  Science, University of Toronto} (\bibinfo{year}{1993}).

\bibitem[{\citenamefont{Ade et~al.}(2013{\natexlab{b}})}]{PlanckXVI}
\bibinfo{author}{\bibfnamefont{P.}~\bibnamefont{Ade}} \bibnamefont{et~al.}
  (\bibinfo{collaboration}{Planck Collaboration})
  (\bibinfo{year}{2013}{\natexlab{b}}), \eprint{astro-ph/1303.5076}.

\bibitem[{\citenamefont{Riess et~al.}(2011)\citenamefont{Riess, Macri,
  Casertano, Lampeitl, Ferguson et~al.}}]{Riess}
\bibinfo{author}{\bibfnamefont{A.~G.} \bibnamefont{Riess}},
  \bibinfo{author}{\bibfnamefont{L.}~\bibnamefont{Macri}},
  \bibinfo{author}{\bibfnamefont{S.}~\bibnamefont{Casertano}},
  \bibinfo{author}{\bibfnamefont{H.}~\bibnamefont{Lampeitl}},
  \bibinfo{author}{\bibfnamefont{H.~C.} \bibnamefont{Ferguson}},
  \bibnamefont{et~al.}, \bibinfo{journal}{Astrophys.J.}
  \textbf{\bibinfo{volume}{730}}, \bibinfo{pages}{119} (\bibinfo{year}{2011}),
  \eprint{1103.2976}.

\end{thebibliography}
\bibliographystyle{apsrev}

\end{document}